\documentclass[a4paper,twocolumn,prl,epsfig]{revtex4}
\usepackage{amssymb}
\usepackage{graphicx}
\usepackage[titles]{tocloft}

\def\v{{\bf v}}
\def\p{{\bf p}}
\def\r{{\bf r}}

\def\q{{\bf q}}
\newcommand{\lt}{\langle r_\alpha r_\beta \rangle_{\rm c}}
\newcommand{\ZZ}{${\mathbb Z}_2$}
\newcommand{\xc}{\langle x^2 \rangle_{\rm c}}
\newcommand{\rab}{{\cal R}_{n,\alpha\beta}}
\newcommand{\iab}{{\cal I}_{n,\alpha\beta}}
\newcommand{\ei}[1]{{\bf e}^{i #1}}
\newcommand{\emi}[1]{{\bf e}^{-i #1}}
\newcommand{\xx}{\mbox{\boldmath$\kappa$}}
\newcommand{\kk}{\mbox{\boldmath$\kappa$}}
\newcommand{\A}{\mbox{\boldmath${\cal A}$}}
\newcommand{\OO}{\mbox{\boldmath$\Omega$}}

\newcommand{\intq}{\int_{\rm BZ} \! d{\bf q} \;}
\newcommand{\da}{\partial_\alpha}
\newcommand{\db}{\partial_\beta}
\renewcommand{\[}{\begin{equation}}
\renewcommand{\]}{\end{equation}}
\def\bea{\begin{eqnarray}}
\def\eea{\end{eqnarray}}
\def\nn{\nonumber\\}
\newcommand{\equ}[1]{Eq.~(\ref{#1})}
\newcommand{\eqs}[2]{Eqs.~(\ref{#1}) and (\ref{#2})}
\def\bra#1{\langle#1\vert}
\def\ket#1{\vert#1\rangle}
\def\ev#1{\langle#1\rangle}

\begin{document}
\title{The Insulating State of Matter: A Geometrical Theory}
\author{Raffaele Resta}                     
%
\affiliation{Dipartimento di Fisica, Universit\`a di Trieste, Italy, \\
and DEMOCRITOS National Simulation Center, IOM-CNR, Trieste, Italy \\ ~ \\
{(\rm Submitted as a ``Colloquium paper'' to Eur. Phys. J. B)}}

\begin{abstract}
\noindent In 1964 W. Kohn published the milestone paper ``Theory of the insulating state'', according to which insulators and metals differ in their {\it ground
state}. Even before the system is excited by any probe, a different organization
of the electrons is present in the ground state and this is the key feature
discriminating between insulators and metals. However, the theory of the
insulating state remained somewhat incomplete until the late 1990s;  this review  addresses the recent developments.
The many-body ground wavefunction of any
insulator is  characterized by means of  geometrical concepts (Berry phase, connection, curvature, Chern number, quantum metric).  Among them, it is the quantum metric which sharply characterizes the insulating state of matter.
The theory deals on a common ground with several kinds of insulators: band insulators, Mott insulators, Anderson insulators,  quantum Hall insulators, Chern and topological insulators
\end{abstract}
\maketitle
\setcounter{tocdepth}{3}
\tableofcontents
\section{Introduction}
\label{intro}

The standard textbook approach to the insulating state of matter is based on band theory. Following Bloch's  theorem \cite{Bloch28} in 1928, the main result is due to Wilson in 1931 \cite{Wilson31}. The single-particle spectrum of a lattice-periodical Hamiltonian is in general gapped, and the electron count determines where the Fermi level lies. If it crosses a band one has a conductor: an applied electric field induces free acceleration of the electrons (at $T=0$ in absence of dissipation). If the Fermi level lies instead in a gap, one has an insulator: in presence of a field
the electronic system polarizes, but no steady-state current flows for
$T \rightarrow 0$. This very successful theory explains the insulating/conducting behavior of most common materials across the periodic table, for which band structure calculations became soon available.

At the root of band theory are two basic assumptions: the electrons are noninteracting (in a mean-field sense), and the solid is crystalline.
By the early 1960s, however, it became clear that there are solids where these
two assumptions are very far from the truth, and where the
insulating behaviour is due to completely different mechanisms. The works of Mott in 1949 \cite{Mott49} and of Anderson in 1958 \cite{Anderson58} opened new avenues in condensed matter physics. In the materials which we now call Mott insulators the insulating behaviour is due to electron correlation \cite{Mott}, while in those called Anderson insulators it is due to lattice disorder \cite{Abrahams}.

In a milestone paper appeared in 1964  Kohn \cite{Kohn64} provided a more comprehensive characterization of the insulating state of matter, which encompasses band insulators, Mott insulators, Anderson insulators, and eventually any kind of insulating material. According to Kohn, the electrons in the insulating state satisfy a many-electron localization condition  \cite{Kohn68}. This kind of localization must be defined in a subtle way given that, for instance, the Hamiltonian eigenstates in a band insulator are obviously {\it not} localized.
According to the original Kohn's formulation, the insulating behaviour arises whenever the ground-state wavefunction of an extended system breaks up into a sum of contributions which are localized in essentially disconnected regions of the many-electron configuration space.

Kohn's theory remained little visited for many years \cite{wos} until the 1990s, when a breakthrough occurred in electronic structure theory: the modern theory of polarization (for historical presentations, see e.g Refs. \cite{rap_a27,rap_a28,rap_a30}). Inspired by the fact that electrical polarization discriminates {\it qualitatively} between insulators and metals, Resta and Sorella \cite{rap107} in 1999 provided a definition of many-electron localization rather different from Kohn's, and deeply rooted in the theory of polarization. Their program was completed soon after by Souza, Wilkens and Martin \cite{Souza00}
(hereafter quoted as SWM), thus providing the foundations of the modern theory of the insulating state. An early review paper appeared in 2002 \cite{rap_a23}.

Several advances occurred afterwards, and the present review illustrates most of them. But the major aim of this work is to provide a thorough formulation of  the modern theory of the insulating state by means of geometrical concepts, which remained somewhat hidden and implicit in the original literature. While the theory of polarization is based on a Berry phase, the theory of the insulating state as presented here is rooted in quantum distance and metric; most of the results previous published in different form will be revisited emphasizing their geometrical content.

\section{Geometry in quantum mechanics} \label{sec:geom}

\begin{figure} 
\centerline{\includegraphics[width=8cm]{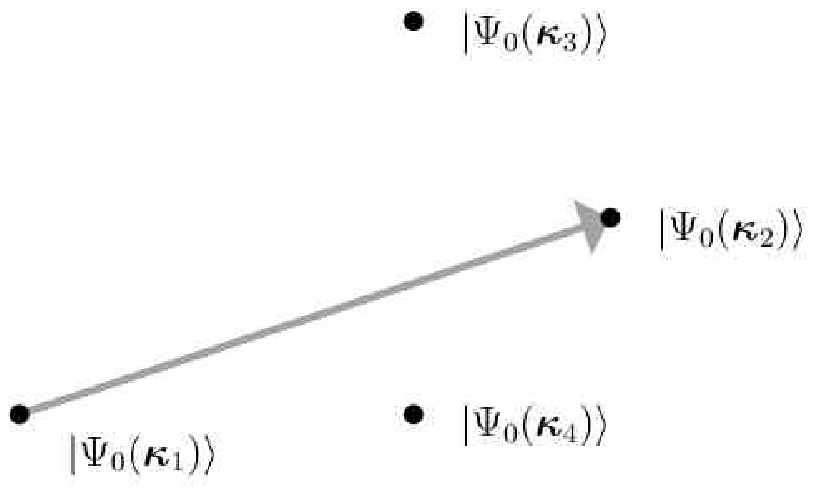}}
\caption{State vectors in the two-dimensional $\kk$-space. The phase difference between two of them is defined as $\emi{\Delta \varphi_{12}} = \frac{\langle \Psi_0(\xx_1) | \Psi_0(\xx_2)  \rangle}{|\langle \Psi_0(\xx_1) | \Psi_0(\xx_2) \rangle|}$, and their distance as $D^2_{12} = 1 - |\langle\Psi_0(\xx_1) | \Psi_0(\xx_2) \rangle|^2$.}
\label{fig:pt1} \end{figure}

Let us assume that a generic time-independent quantum Hamiltonian has a parametric dependence. The Schr\"odinger equation is
\[ H(\kk) \ket{\Psi(\kk)} =  E(\kk) \ket{\Psi(\kk)} , \]
where the $d$-dimensional real parameter $\kk$ is defined in a suitable domain
of ${\mathbb R}^d$: a 2d  $\kk$  has been chosen for display in Fig.~\ref{fig:pt1}. In this Section we start discussing the most general case, and therefore for the time being we do not specify which quantum system is described by this Hamiltonian, nor what the physical meaning of the parameter $\kk$ is. Later on, $\ket{\Psi(\kk)}$ will be identified with a many-electron wavefunction. The state vectors  $\ket{\Psi(\kk)}$ are all supposed to be normalized and to reside in the same Hilbert space: this amounts to saying that the wavefunctions are supposed to obey $\kk$-independent boundary conditions. 
We focus on the ground state $\ket{\Psi_0(\kk)}$, and we assume it  to be nondegenerate for $\kk$ in some domain of ${\mathbb R}^d$.

Any quantum mechanical state vector is arbitrary by a constant phase factor. Here we refer to choosing this phase as to the choice of the gauge. All measurable quantities (e.g. expectation values) are obviously gauge-invariant, but the reverse is also true: all gauge-invariant properties are---at least in principle---measurable.

It is expedient to define the ground-state projector (a.k.a. density matrix) and its complement, i.e.  \[ \hat{P}(\kk) = \ket{\Psi_0(\kk)} \bra{\Psi_0(\kk)} ; \quad \hat{Q}(\kk) = \hat{1} - \hat{P}(\kk) . \label{proj} \] Both
$\hat{P}(\kk)$ and $\hat{Q}(\kk)$ are obviously gauge-invariant.

\subsection{Phases and distances}

We define the {\it phase difference} between the ground eigenstates at
two different $\xx$ points in the most natural way: \[  \emi{\Delta \varphi_{12}} = \frac{\langle \Psi_0(\xx_1) | \Psi_0(\xx_2)  \rangle
}{|\langle \Psi_0(\xx_1) | \Psi_0(\xx_2) \rangle|} \; ; \label{pancha1} \]  
\[ \Delta \varphi_{12} = -\mbox{ Im log } \langle \Psi_0(\xx_1) | \Psi_0(\xx_2) 
\rangle \label{pancha2} \; . \] For any given choice of the two states,
\eqs{pancha1}{pancha2} provide a $\Delta \varphi_{12}$ which is unique modulo
$2 \pi$, except in the very special case that the states are orthogonal.
However, it is also clear that such $\Delta \varphi_{12}$ is gauge-dependent and {\it cannot} have, by itself, any physical meaning. 

The distance between quantum states has been defined by Bures \cite{Bures69} as: \[ D^2_{12} = 1 - |\langle\Psi_0(\xx_1) | \Psi_0(\xx_2) \rangle|^2 . \label{bures} \] Such distance fulfills the familiar axioms from calculus textbooks; it vanishes when the two states physically coincide (i.e. independently of the phase factor), and is maximum (equal to one) when the states are orthogonal. At variance with $\Delta \varphi_{12}$ defined above, the Bures distance is  gauge-invariant, and can be explicitly expressed in terms of ground-state projectors \[ D^2_{12} = 1 - \mbox{Tr } \{ \hat{P}(\kk_1) \hat{P}(\kk_2) \} , \] where ``Tr'' is the trace over the Hilbert space.

\subsection{Berry phase}

\begin{figure} 
\centerline{\includegraphics[width=8cm]{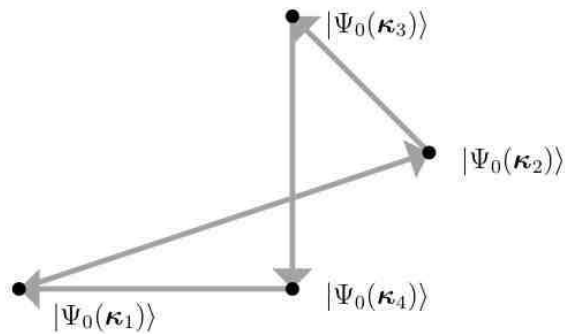}}
\caption{A closed path joining four states in $\kk$-space.}
\label{fig:pt2} \end{figure}

We have already observed that the phase difference $\Delta \varphi_{12}$ between any two states is gauge-dependent and cannot have any physical meaning. Matters are quite different when we consider the
{\it total} phase difference along a closed path which joins several points in a
given order, as shown in Fig.~\ref{fig:pt2}: \bea  \gamma &=&
\Delta \varphi_{12} +  \Delta \varphi_{23} +  \Delta \varphi_{34} +  \Delta
\varphi_{41} \nn &=& - \mbox{ Im log } \langle \Psi_0(\xx_1) |
\Psi_0(\xx_2) \rangle \langle \Psi_0(\xx_2) | \Psi_0(\xx_3) \rangle \times \nn &\times& \langle
\Psi_0(\xx_3) | \Psi_0(\xx_4) \rangle \langle \Psi_0(\xx_4) | \Psi_0(\xx_1) \rangle .
\label{discre1} \eea It is now clear that all the gauge-arbitrary
phases cancel in pairs, such as to make the overall phase $\gamma$ a
gauge--invariant quantity. The above simple--minded algebra leads to a result of
overwhelming physical importance: in fact, a
gauge--invariant quantity is potentially a physical observable. In essence, this is the revolutionary message of Berry's celebrated paper, appeared in 1984 \cite{Berry84,Berry}.

\begin{figure} 
\centerline{\includegraphics[width=6cm]{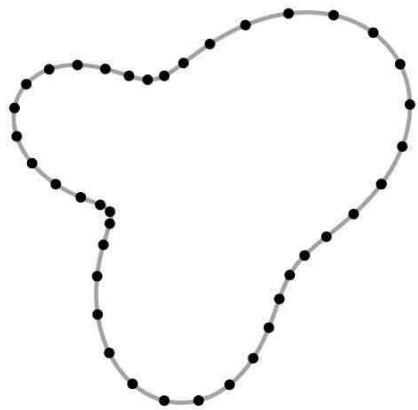}}
\caption{A smooth closed curve $C$ in $\kk$-space, and its discretization.}
\label{fig:pt3} \end{figure}

Next we consider a smooth closed curve $C$ in the
parameter domain, such as in Fig.~\ref{fig:pt3}, and we discretize it with
a set of points on it. Using \equ{pancha1}, we write the phase difference
between any two contiguous points as \[  \emi{\Delta
\varphi} = \frac{\langle \Psi_0(\xx) | \Psi_0(\xx \!+\! \Delta \xx)  \rangle
}{|\langle \Psi_0(\xx) | \Psi_0(\xx \!+\! \Delta \xx) \rangle|} . \label{delta}
\] If we further assume that the gauge is so chosen that the phase varies in
a {\it differentiable} way along the path, then from \equ{delta} we get to
leading order in $\Delta \xx$: \[ -i \Delta \varphi \simeq \langle \Psi_0(\xx)
| \nabla_{\kk} \Psi_0(\xx)  \rangle \cdot \Delta \xx .  \] In the limiting case of a
set of points which becomes dense on the continuous path, the total phase
difference $\gamma$  converges to a circuit
integral: \[ \gamma = \sum_{s=1}^{M} \Delta \varphi_{s,s+1}  \;
\longrightarrow \; \oint_{C} \A(\kk) \cdot d \kk , \label{sum} \] where $\A(\kk)$ is called the Berry connection: \[ \A(\kk) = i  \, \langle
\Psi_0(\xx) | \nabla_{\kk} \Psi_0(\xx)  \rangle . \label{berry1} \] Since the
state vectors are assumed to be normalized at any $\xx$, the connection  is {\it real}; we can therefore equivalently write  \[ \A(\kk)  = - \mbox{Im } \langle \Psi_0(\xx) | \nabla_{\kk} \Psi_0(\xx)  \rangle  . \label{berry2} \] 

Dealing with the disparate manifestations of the Berry phase which occur in molecular and condensed matter phenomena is clearly beyond the scope of the present review; we quote Refs. \cite{Berry,Thouless,Geometric,rap_a20,Xiao10} out of many possible references.

\subsection{Connection and curvature}

The loop integral of the Berry connection (i.e. the Berry phase $\gamma$) is non trivial in two cases: either the curl of $\A(\kk)$ is nonzero, or the curl is zero but the curve $C$
is not in a simply connected domain. In the former case, we can invoke Stokes theorem; the formulation is very simple when $\kk$ is a 3d parameter. If $C$ is the boundary of a surface $\Sigma$ (i.e. $C \equiv \partial \Sigma$), and the curl of $\A(\kk)$ is regular on $\Sigma$, then
Stokes' theorem reads \[ \gamma = \oint_{\partial \Sigma}  \A(\kk) \cdot d \kk  = \int_\Sigma \OO(\kk) \cdot {\bf n}  \; d \sigma , \label{sto} \]
where $\OO$ is the Berry curvature, defined as \bea \OO(\kk) &=& \nabla_{\kk} \times \A(\kk)  =  - \mbox{Im } \langle \nabla_{\kk}
\Psi_0(\xx) | \times | \nabla_{\kk} \Psi_0(\xx)  \rangle \nn &=& i \langle \nabla_{\kk} 
\Psi_0(\xx) | \times | \nabla_{\kk} \Psi_0(\xx)  \rangle , \label{curva} \eea with the usual meaning of the cross product between three-component bra and ket states. Equation (\ref{sto}) may be spelled out by saying that the curvature is the Berry phase per unit area of $\Sigma$.

For $d \neq 3$ the Berry curvature is conveniently written as the $d \times d$ antisymmetric matrix \[ \Omega_{\alpha\beta}(\kk) = -2 \, \mbox{Im } \ev{ \da \Psi_0(\kk) | \db \Psi_0(\kk)} ; \label{curva3} \] Greek subscripts are Cartesian coordinates throughout, and $\da = \partial / \partial \kappa_\alpha$. The Stokes theorem can still be applied, generalizing \equ{sto} to \[ \gamma = \frac{1}{2} \int_\Sigma d \kappa^\alpha \wedge d \kappa^\beta \; \Omega_{\alpha\beta}(\kk) . \label{stog} \]

The Berry connection is also known as ``gauge potential", and the Berry curvature as ``gauge field'' \cite{Geometric}. It is worth pointing out that the former is gauge-dependent, while the latter is gauge-invariant and therefore corresponds in general to a measurable quantity, even before any integration. The two quantities play (in $\kk$-space) a similar role as the vector potential and the magnetic field in elementary magnetostatics: ${\bf A}(\r)$ is gauge-dependent, nonmeasurable; ${\bf B}(\r) = \nabla_{\r} \times {\bf A}(\r)$ is gauge-invariant, measurable.

The Berry phase $\gamma$, defined as the integral over a closed curve $C$ of the connection, is gauge invariant only modulo $2\pi$. This indeterminacy is resolved by \eqs{sto}{stog} whenever the curve $C$ is the boundary $\partial \Sigma$ of a surface $\Sigma$ where the curvature is regular. In fact, the curvature is gauge-invariant and has no modulo $2\pi$ indeterminacy.

\subsection{Chern number} \label{sec:chern}

The rhs of \eqs{sto}{stog} is the flux of the Berry curvature on the surface 
$\Sigma$; such flux remains meaningful even on a closed surface (e.g. a sphere or a torus), in which case $\partial \Sigma$ is the empty set. The key result is that
such an integral is quantized. Here we limit ourselves to 3d where \[ \int_\Sigma \OO(\kk) \cdot {\bf n}  \; d \sigma = 2\pi C_1; \label{chern} \]  $C_1$ is an integer, called Chern number of the first class.

The proof is based on a similar algebra as for Dirac's theory of the magnetic monopole \cite{Thouless,SakuraiM}, and is also closely related to the Gauss-Bonnet theorem in differential geometry. 

The curvature is regular (and divergence-free) on the closed surface $\Sigma$; the lhs of \equ{chern} is the flux of $\OO(\kk)$ across $\Sigma$. The integrand $\OO(\kk)$ is the curl of the connection $\A(\kk)$; the latter in general {\it cannot} be defined as a single-valued function globally on $\Sigma$, but only on {\it patches} of it \cite{Thouless,SakuraiM}.
To fix the ideas, suppose that $\OO(\kk)$ is singular at $\kk=0$, and that $\Sigma$ is a spherical surface centered at the origin. We cut this surface at the equator $\kappa_z=0$ and we consider the flux across the two open surfaces: \[
\int_\Sigma \OO(\kk) \cdot {\bf n}  \; d \sigma = \int_{\Sigma_+} \OO(\kk) \cdot {\bf n}  \; d \sigma + \int_{\Sigma_-} \OO(\kk) \cdot {\bf n}  \; d \sigma  .\] We notice that $\partial \Sigma_+ = \partial \Sigma_- = C$, but the surface normals ${\bf n}$ have opposite orientations. From Stokes theorem, \equ{sto}, we get: \[ \int_{\Sigma_\pm} \OO(\kk) \cdot {\bf n}  \; d \sigma = \pm \oint_{C} \A_\pm(\kk) \cdot d \kk \] \[  \int_{\Sigma} \OO(\kk) \cdot {\bf n}  \; d \sigma =  \oint_{C} \A_+(\kk) \cdot d \kk - \oint_{C} \A_-(\kk) \cdot d \kk \; . \label{dirac} \] The two upper and lower Berry connections $\A_\pm(\kk)$ may only differ by a gauge transformation; the rhs of \equ{dirac} is the difference of two Berry phases on the same path and is necessarily a multiple of $2\pi$. This concludes the proof of \equ{chern}.

We emphasize that the Chern number is a robust topological invariant of the wavefunction, and is at the origin of observable effects. The
Chern number made its first appearance in electronic structure in 1982, in the famous TKNN paper about the quantum Hall effect \cite{Thouless82}. Nowadays more complex topological invariants are in fashion, and they characterize a completely novel class of insulators, called ``topological insulators'' \cite{Kane05,Sheng06,Zhang08,Chen09,Moore09,Qi10,Hasan10}.

\subsection{Metric}

Starting from  \equ{bures}, the infinitesimal distance is
\[ D^2_{\kk,\kk + d\kk} = \sum_{\alpha,\beta=1}^d  g_{\alpha\beta}(\xx) d
\kappa_\alpha d \kappa_\beta , \] where the metric tensor is easily shown to be
\bea g_{\alpha\beta}(\xx) &=& \mbox{Re } \ev{ \da \Psi_0(\kk) | \db \Psi_0(\kk)} 
\nn &-&  \ev{ \da \Psi_0(\kk) |  \Psi_0(\kk)} \ev{ \Psi_0(\kk) | \db \Psi_0(\kk)}
\nn &=& \mbox{Re } \ev{ \da \Psi_0(\kk) | \hat{Q}(\kk)| \db \Psi_0(\kk)} ; \label{metric} \eea the projector $\hat{Q}(\kk)$ is the same as defined in \equ{proj}.
This quantum metric tensor was first proposed by Provost and Vallee in 1980 \cite{Provost80}.

At this point we may compare \equ{metric} to \equ{curva3}, noticing that the insertion of $\hat{Q}(\kk)$ is irrelevant in the latter, i.e. \[ \Omega_{\alpha\beta}(\kk) = -2 \, \mbox{Im } \ev{ \da \Psi_0(\kk) | \hat{Q}(\kk) | \db \Psi_0(\kk)} . \label{curva4} \] 
It is therefore clear that $g_{\alpha\beta}$ and $\Omega_{\alpha\beta}$ are, apart for a trivial $-2$ factor, the real (symmetric) and the imaginary (antisymmetric) parts of the same tensor, which we are going to call $\eta_{\alpha\beta}$ in the following: \[ \eta_{\alpha\beta}(\kk) =  \ev{ \da \Psi_0(\kk) | \hat{Q}(\kk) | \db \Psi_0(\kk)} . \label{eta} \] The curvature, the metric, and hence the full tensor $\eta_{\alpha\beta}$ are all gauge-invariant quantities. A compact equivalent expression is \[ \eta_{\alpha\beta}(\kk) = \mbox{Tr } \{ \da \hat{P}(\kk) \hat{Q}(\kk) \db \hat{P}(\kk) \} , \label{eta2} \] manifestly gauge-invariant and Hermitian.

\subsection{Sum over states}

The $\kk$-derivatives entering many of the previous equations, e.g. \equ{eta},
can be expressed starting from  perturbation theory:
\bea && \ket{\Psi_0(\kk + \Delta \kk)} - \ket{\Psi_0(\kk)} \label{pertu}  \\ &\simeq&  {\sum_{n \neq 0}}' \ket{\Psi_n(\kk)} \frac{\ev{\Psi_n(\kk) | \, [ \, H(\kk + \Delta \kk) - H(\kk) \, ] \, | \Psi_0(\kk)}}{E_0(\kk) - E_n(\kk)} ; \nonumber  \eea
 \[ \ket{\da \Psi_0(\kk)} = {\sum_{n \neq 0}}' \ket{\Psi_n(\kk)} \frac{\ev{\Psi_n(\kk) | \da H(\kk) | \Psi_0(\kk)}}{E_0(\kk) - E_n(\kk)} .  \label{pertux} \] Thiese seemingly obvious and innocent formulae need some caveat. It is clear that inserting \equ{pertux} into the Berry connection, \eqs{berry1}{berry2}, we get a vanishing result for any $\kk$. This happens because the simple expression of \equ{pertu} corresponds to a very specific gauge choice (called the parallel-transport gauge \cite{rap_a20}); multiplying the rhs by a $\kk$-dependent phase factor is legitimate, and must not modify any physical result, while the Berry connection is instead affected. Nonetheless, since our 
$\eta_{\alpha\beta}(\kk)$ is a gauge-invariant quantity, we may safely evaluate it in any gauge, including the parallel-transport gauge, implicit in \equ{pertux}. The result is
\bea \eta_{\alpha\beta}(\kk) && \label{pertu2} \\ = {\sum_{n \neq 0}}'  && \frac{\ev{\Psi_0(\kk) | \da H(\kk) | \Psi_n(\kk)} \ev{\Psi_n(\kk) | \db H(\kk) | \Psi_0(\kk)}}{[ E_0(\kk) - E_n(\kk) ]^2} .  \nonumber  \eea This expression shows explicitly that both the curvature and the metric are ill defined and singular wherever the ground state is degenerate with the first excited state. Indeed, this is the main reason why the domain may happen not to be simply connected.

\section{Many-electron wavefunction}

In the previous Section we have discussed some geometrical aspects of quantum mechanics, remaining at a rather abstract level, and without addressing either specific physical systems, or measurable quantities. From now on, instead, we address a many-electron system, whose most general Hamiltonian we write, in the Schr\"odinger representation and in Gaussian units, as \[ \hat{H}(\kk) = \frac{1}{2 m_e} \sum_{i=1}^N | \p_i + \frac{e}{c} {\bf A}(\r_i) + \hbar \kk |^2 + \hat{V} , \label{sch} \] where $m_e$ and $- e$ are the electron mass and charge, respectively ($e>0$).
Equation (\ref{sch}) is exact in the nonrelativistic, infinite-nuclear-mass limit. In \equ{sch} $\bf{A}(\r)$ is a vector potential of magnetic origin and the potential $\hat{V}$ includes one-body and two-body (electron-electron) contributions. The 3d parameter $\kk$, having the dimensions of an inverse length, can be regarded as a pure gauge, constant in space, which does not affect the fields (electric and magnetic) and therefore is irrelevant in classical mechanics. Matters are different in quantum mechanics, where the {\it potentials} (vector and scalar) are the most fundamental quantities, as we know well since the Bohm-Aharonov paper in 1959 \cite{AeB,Feynman2}.
In our case the parameter $\kk$, called ``flux'' or ``twist'', affects indeed the eigenvalues and eigenvectors depending on the chosen boundary conditions.

\subsection{Boundary conditions}

Two very different kinds of boundary conditions occur in electronic structure theory: ``open'' boundary conditions (OBCs) and periodic boundary conditions (PBCs). The former are appropriate to molecular physics, and require that the many-electron wavefunction of a bound state is square-integrable over the whole coordinate space ${\mathbb R}^{3N}$. PBCs are instead appropriate for extended systems, either crystalline or disordered, either independent-electron or correlated. For the sake of simplicity here we adopt PBCs over a cubic box of side $L$, meaning  that the eigenstates of \equ{sch} at any given $\kk$ are
Born-von-K\`arm\`an periodic in the cubic box over each electron
coordinate independently; their Cartesian components $r_{i,\alpha}$ are then
equivalent to the angles $2 \pi r_{i,\alpha}/L)$. The potential $\hat{V}$
enjoys the same periodicity, which implies that the microscopic electric field averages to
zero over the sample. The eigenfunctions are normalized in the hypercube of volume $L^{3N}$.

As previously observed, the choice of boundary conditions (either OBCs or PBCs) corresponds to choosing the Hilbert space, which in turn affects profoundly the geometrical properties discussed in Sec.~\ref{sec:geom}.

\subsubsection{Open boundary conditions}

The case of OBCs is by far the simplest. We write for the sake of simplicity the ground state of \equ{sch} at $\kk = 0$ as $\ket{\Psi_0} \equiv \ket{\Psi_0(0)}$, and we define the many-body position operator as \[ \hat{\r} = \sum_{i=1}^{N} \r_i . \label{pos} \] It is straightforward to verify that the state $\emi{\kk \cdot  \hat{\r}}
\ket{\Psi_0}$ fulfills both the Schr\"odinger equation at the given $\kk$ and the boundary conditions with a $\kk$-independent eigenvalue. In jargon we say that the $\kk$-dependence is easily ``gauged away'' within OBCs; we anticipate that matters are quite different within PBCs.

The state $\emi{\kk \cdot  \hat{\r}} \ket{\Psi_0}$ coincides therefore with the ground eigenstate $\ket{\Psi_0(\kk)}$ of the twisted Hamiltonian, \equ{sch}. It is legitimate to multiply
this eigenvector by any $\kk$-dependent (and position-independent) phase factor;  our choice is then \[ \ket{\Psi_0(\kk)} = \emi{\kk \cdot  ( \hat{\r} - {\bf d} )}
\ket{\Psi_0} , \] where ${\bf d} = \ev{\Psi_0 | \hat{\r} | \Psi_0}$ is the electronic dipole of the molecular system. It follows that the $\kk$-derivative needed in our geometrical quantities (curvature and metric) is 
\[ \ket{\nabla_{\kk} \Psi_0} = -i ( \hat{\r} - {\bf d} )
\ket{\Psi_0}  = -i \hat{Q}(0) \, \hat{\r} \ket{\Psi_0} , \label{Qr2} \] Exploiting the idempotency of the projector, the tensor $\eta_{\alpha\beta}$, evaluated at $\kk=0$, is
\bea \eta_{\alpha\beta}(0) &=&  \ev{\Psi_0 | \hat{r}_\alpha \hat{Q}(0)  \hat{r}_\beta | \Psi_0} \nn &=& \ev{\Psi_0 | \hat{r}_\alpha  \hat{r}_\beta | \Psi_0} -
\ev{\Psi_0 | \hat{r}_\alpha  | \Psi_0}\ev{\Psi_0 |  \hat{r}_\beta | \Psi_0}
\label{eta3} . \eea This is clearly a real symmetric tensor, hence the Berry curvature vanishes within OBCs, and the metric $g_{\alpha\beta}(0)$ coincides with $\eta_{\alpha\beta}(0)$. It will be shown that within PBCs, instead, the tensor $\eta_{\alpha\beta}(0)$ may acquire a nonvanishing imaginary part.

Clearly, the metric tensor in \equ{eta3} is the cumulant second moment of the position operator, or equivalently the ground state fluctuation of the dipole of the molecular system; this quantity is extensive (scales like $N$ in macroscopically homogenous systems). We anticipate that $\eta_{\alpha\beta}(0)/N$ discriminates, in the large $N$-limit, between insulators and metals.

\subsubsection{Periodic boundary conditions}

First of all, some preliminary observations about the position operator are in order. The simple multiplicative operator $\hat{\r}$, as defined in \equ{pos}, is ``forbidden'' within PBCs: in fact it maps any periodic wavefunction $\ket{\Psi}$
into the {\it nonperiodic} function $\hat{\r} \ket{\Psi}$. In other words, it maps a function which belongs to the Hilbert space to a function outside of it \cite{rap100}. Incidentally, this is the main reason why the polarization problem has
remained unsolved until the early 1990s (for historical presentations, see e.g Refs. \cite{rap_a27,rap_a28,rap_a30}). In the present context, formulae like \equ{eta3} are ill defined and absurd within PBCs.

Suppose that $\ket{\Psi_0}$ is the ground eigenstate of \equ{sch} at $\kk=0$. Then the simple state $\emi{\kk \cdot  \hat{\r}}
\ket{\Psi_0}$ fulfills the Schr\"odinger equation but {\it not}---for a general $\kk$---PBCs. It is therefore not an eigenstate, except for the discrete $\kk$ values
\[ \kk_{m_1 m_2 m_3} = \frac{2\pi}{L} (m_1 {\bf e}_1 + m_2 {\bf e}_2 + m_3 {\bf e}_3) , \label{kdiscre} \] where $m_\alpha \in {\mathbb Z}$, and ${\bf e}_\alpha$ are the Cartesian versors. For fractional values of $\kk$---i.e. for values different from those in \equ{kdiscre}---the $\kk$-dependence of the eigenvalues and eigenvectors of \equ{sch} is nontrivial.

If $\ket{\Psi_0(\kk)}$ is the genuine ground eigenstate of \equ{sch} within PBCs, then the auxiliary function $\ket{\tilde{\Psi}_0(\kk)} = \ei{\kk \cdot  \hat{\r}} \ket{\Psi_0(\kk)}$ is a solution of $\hat{H}(0)$, but fulfills quasi-periodic boundary conditions: at any two opposite faces of the cube the wavefunction differs by a $\kk$-dependent phase factor. In other words the problem can be formulated in two equivalent ways: either the Hamiltonian is $\kk$-dependent, as in \equ{sch}, and the boundary conditions are $\kk$-independent; or the Hamiltonian is $\kk$-independent but the boundary conditions are ``twisted'' in a $\kk$-dependent way.

\begin{figure} 
\centerline{\includegraphics[width=4.5cm]{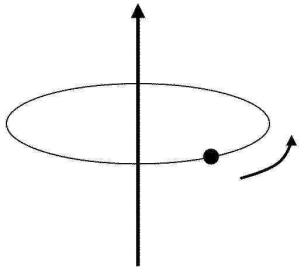}}
\caption{The electron motion is confined to a circular rail. A constant vector potential along the rail, as in \protect\equ{sch1}, corresponds to vanishing fields (electric and magnetic), yet the spectrum {\it depends} on the ``inaccessible flux'' piercing the surface encircled by the rail. }
\label{fig:flux} \end{figure}

We briefly switch to a 1d formulation in order to explain the role of $\kk$, and even the alternative semantics (``flux'' instead of ``twist''). Setting the magnetic vector potential to zero, the 1d version of \equ{sch} is  \[ \hat{H}(\kappa) = \frac{1}{2 m_e} \sum_{i=1}^N ( p_i  + \hbar \kappa )^2 + \hat{V} , \label{sch1} \]
and we require the wavefunction to be periodical with period $L$ over each electron coordinate independently. Alternatively, we may regard this problem as if
the electrons were confined to a circular rail of circumference $L$ (see Fig. \ref{fig:flux}). There is no fields (electric or magnetic)  on the rail, but the Hamiltonian is the same as if a constant vector potential of intensity $A=c \hbar \kappa / e$ were present along the rail. This corresponds to a magnetic  flux $\Phi =c \hbar \kappa L / e$ piercing the surface encircled by the rail, in a region {\it not visited} by the electronic system; it has been appropriately called by some authors ``inaccessible flux''. This flux affects nontrivially the eigenvalues and eigenvectors, 
as it is easily verified in the trivial case $\hat{V} \equiv 0$. The inaccessible flux has therefore observable effects; a similar algebra in a different context leads indeed to the Bohm-Aharonov effect \cite{AeB,Feynman2}. When $\kappa$ assumes one of the values in \equ{kdiscre}, then the flux $\Phi$ is an integer multiple of the elementary flux quantum $\Phi_0 = 2\pi \hbar c /e = h c / e$  ($\Phi_0 = h/e$ in SI units). We stress that  only the fractional part of the flux affects the results in a nontrivial way.

It has already been observed that within PBCs the tensor $\eta_{\alpha\beta}(0)$ cannot be simplified---as e.g. in \equ{eta3}---and must be therefore addressed in its original form, \equ{eta}, by actually evaluating the $\kk$ derivatives. As within OBCs, even within PBCs these tensors are extensive. In the thermodynamic limit (i.e $N \rightarrow \infty$, keeping $N/L^3$ constant), the well defined quantity
is $\eta_{\alpha\beta}(0)/N$.

Finally, we observe that for time-reversal symmetric systems $\eta_{\alpha\beta}(0)$ is real symmetric, and coincides therefore with the metric tensor $g_{\alpha\beta}(0)$. 

\section{The localization tensor}

In the milestone paper, appeared in 1964 under the title ``Theory
of the insulating state'' \cite{Kohn64,Kohn68}, W. Kohn addressed the
insulating character of a solid from a very general viewpoint. Kohn focusses on the electronic {\it ground state}
as a whole. His outstanding message is that ``insulating behavior does
not appear to depend on the notion of a filled band but rather
reflects a certain type of organization of the electrons''.
Furthermore, the electrons ``so organize themselves as to satisfy a
many electron localization condition'' \cite{Kohn68}. Such localization,
 however, must be defined in a subtle way.  For instance,
the (single-particle) Hamiltonian eigenstates of a band insulator are
definitely {\it not} localized. 

The modern formulations of the theory of the insulating state are all based on a
localization tensor (squared localization length in 1d), first introduced
by Resta and Sorella in 1999 \cite{rap107}, as the main tool to address the ``organization of the electrons'' in the ground state. This work was followed soon afterwards by SWM \cite{Souza00}, where additional results are found, most notably relating localization to conductivity. Since then, most authors (including the present one) have adopted the notation $\lt$ for the localization tensor, where ``c'' stays for cumulant. The formulation within OBCs, using the language and the notations familiar in quantum chemistry, dates since 2006 \cite{rap132}.

The tensor $\lt$ has the dimensions of a squared length; it is an intensive quantity that characterizes the ground-state many-body wavefunction as a whole. Its key virtue is that it discriminates between insulators and metals: it is finite in the former case and divergent (in the large-system limit) in the latter. This is the main message of the present work (and of the modern theory of the insulating state): we are going to prove it below, Sect. \ref{sec:loca}

Several definitions of the localization tensor have been given in the literature, all of them equivalent. Here we give a novel one, deeply rooted in the geometrical concepts discussed so far: the localization tensor is the intensive quantity
\[ \lt = \eta_{\alpha\beta}(0)/N , \] where the thermodynamic limit is understood.
A glance at the OBCs expression, \equ{eta3}, explains the reason for the notation, which we adopt within PBCs as well.

Until 2005 the theory of the insulating state implicitly addressed time-reversal invariant systems only, where the $\eta$ tensor is real symmetric, and coincides
with the metric: in such systems therefore \[ \lt = g_{\alpha\beta}(0)/N . \] It was found in 2005 \cite{rap127} that---in absence of time-reversal symmetry and within PBCs---the tensor $\lt$ is naturally endowed with an antisymmetric imaginary part, whose physical meaning is also outstanding. This is discussed in Sects. \ref{sec:cins} and \ref{sec:qh}.

\subsection{Discretized  formulae}

While practical implementations within OBCs invariably use \equ{eta3}, those within PBCs require a discretization in $\kk$-space in order to evaluate the derivatives appearing in \equ{eta}. A key feature is that any discretization must be
{\it numerically} gauge invariant. Here we are going to present a novel discretized formula, based on \equ{eta2} at $\kk=0$. 

By definition we have \[ \da \hat{P}(0) = \lim_{k \rightarrow 0}
\frac{1}{k} [  \, \hat{P}(k {\bf e}_\alpha) - \hat{P}(0) \, ]  , \] hence for small $k$
\bea \eta_{\alpha\beta}(0) &\simeq& \frac{1}{k^2} \mbox{Tr } \{  [  \, \hat{P}(k {\bf e}_\alpha) - \hat{P}(0) \, ] \hat{Q}(0) [  \, \hat{P}(k {\bf e}_\beta) - \hat{P}(0) \, ] \}
\nn &=& \frac{1}{k^2} \mbox{Tr } \{  \hat{P}(k {\bf e}_\alpha)  \hat{Q}(0)  \hat{P}(k {\bf e}_\beta) \} \\ &=&  \frac{1}{k^2} \ev{\Psi_0(k {\bf e}_\alpha) | \Psi_0(k {\bf e}_\beta)} \ev{\Psi_0(k {\bf e}_\beta) | \Psi_0(k {\bf e}_\alpha)}  \label{three}
\\ &-&  \frac{1}{k^2} \ev{\Psi_0(k {\bf e}_\alpha) | \Psi_0} \ev{\Psi_0 | \Psi_0(k {\bf e}_\beta)}  \ev{\Psi_0(k {\bf e}_\beta) | \Psi_0(k {\bf e}_\alpha)} . \nonumber \eea
The gauge invariance is pespicuous: all the arbitrary phase factors cancel in pairs.

Equation (\ref{three})  is still a three-point formula; it is transformed to a single-point one using a key result from Ref. \cite{rap107}. 
Whenever $k$ is equal to $\frac{2\pi}{L}$ times an integer, the wavefunction $\emi{k  \hat{r}_\alpha} \ket{\Psi_0(0)}$ obeys PBCs---see also \equ{kdiscre}---and is an eigenstate of $\hat{H}(k {\bf e}_\alpha)$ with eigenvalue $E_0$. For insulating systems the ground state is nondegenerate even in the thermodynamic limit, hence \[ \ket{\Psi_0(\frac{2 \pi}{L} {\bf e}_\alpha)}  = \emi{2 \pi \hat{r}_\alpha / L} \ket{\Psi_0} , \label{twist} \] apart possibly by a phase factor, irrelevant here. Replacing  $k$ with $\frac{2 \pi}{L}$ for large $L$ in \equ{three}, and using \equ{twist} we get the sought for formula, where a single diagonalization of \equ{sch} at $\kk=0$ is enough to obtain both the real and imaginary part of the localization tensor $\lt$. Notice that the first line of \equ{three} is real.

The formulae previous reported in the literature are inspired by the Berry-phase formulae---see \equ{discre1}---and based on logarithms. To retrieve them, it is enough to use $\log (1+\varepsilon) \simeq \epsilon$ in \equ{three}: \[ \eta_{\alpha\beta}(0) \simeq \frac{1}{k^2} \log  \frac{ \ev{\Psi_0(k {\bf e}_\alpha) | \Psi_0(k {\bf e}_\beta)} }{ \ev{\Psi_0(k {\bf e}_\alpha) | \Psi_0 } \ev{ \Psi_0 |\Psi_0(k {\bf e}_\beta)} } , \label{loga} \] which provides $\lt$  identical in form to Eq. (86) in Ref. \cite{rap_a23}; for $\alpha=\beta$  the formula is also identical to the archetypical one, proposed by Resta and Sorella in 1999 \cite{rap107}. For the real part of $\lt$ the logarithm formula, \equ{loga}, has some advantages over \equ{three}. For the imaginary part, instead, the second line of \equ{three} provides a much better result, since it is {\it not} affected by any modulo $2\pi$
indeterminacy. Getting rid of such indeterminacy, in fact, is essential when dealing with Chern numbers (Sects. \ref{sec:cins} and \ref{sec:qh}).

\subsection{Sum over states again}

We express $\lt$  using the sum-over-states formula, \equ{pertu2} : \[ \lt = \frac{1}{N} {\sum_{n \neq 0}}' \frac{\ev{\Psi_0 | \da \hat{H}(0) | \Psi_n} \ev{\Psi_n | \db \hat{H}(0) | \Psi_0}}{( E_0 - E_n )^2} . \label{pertu3} \] 
The $\kk$-derivative of the Hamiltonian of \equ{sch} is \[ \nabla_{\kk} \hat{H}(0) = 
 \frac{\hbar}{m_e} \sum_{i=1}^N \left[ \, \p_i + \frac{e}{c} {\bf A}(\r_i) \, \right] , \] where the rhs is nothing else than $\hbar$ times the velocity operator $\hat{\v}$; \equ{pertu3} becomes then \bea \lt  &=& \frac{1}{\hbar^2 N} {\sum_{n \neq 0}}' \frac{\ev{\Psi_0 | \hat{v}_\alpha | \Psi_n} \ev{\Psi_n | \hat{v}_\beta | \Psi_0}}{( E_0 - E_n )^2}  \nn &=& \frac{1}{N} {\sum_{n \neq 0}}' \frac{\ev{\Psi_0 | \hat{v}_\alpha | \Psi_n} \ev{\Psi_n | \hat{v}_\beta | \Psi_0}}{\omega_{0n}^2}  , \label{pertu4} \eea where $\omega_{0n} = (E_n - E_0)/\hbar$.
 
 The velocity operator is also commonly expressed as 
 $\hat{\v} = i [\hat{H}(0), \hat{\r} ] /\hbar$, but it is worth emphasizing that while the position $\hat{\r}$ is well defined within OBCs and ill defined within PBCs, the velocity $\hat{\v}$ is well defined in both cases. 
 
The basic sum-over states formula, \equ{pertu4}, applies therefore to both OBCs and PBCs. In general, it is not very useful on practical grounds, since it would require the evaluation of slowly convergent sums. Nonetheless, \equ{pertu4} is instead essential to gather understanding into the physical meaning of
$\eta_{\alpha\beta}$, as will be shown below.

\section{Conductivity}

\subsection{Linear response}

So far, we have only discussed ground-state properties of our $N$-electron
system.  Suppose now that it is subject to a small time-dependent
perturbation contributing to the Hamiltonian the term: \[ \delta \hat{H}(t) =
\frac{1}{2\pi} \int_{-\infty}^\infty \!\!\! d \omega \; f(\omega) \frac12
(\hat{A} \emi{\omega t} + \hat{A}^\dagger \ei{\omega t}) , \] where $\hat{A}$
determines the ``shape'' of the perturbation and $f$ its amplitude. In order
to get an Hermitian $\delta \hat{H}$, we assume $f(\omega) = f(-\omega)$. We
wish to measure the response to such perturbation by means of the expectation
value of some observable $\hat{B}$, i.e.: \[ \delta \langle \hat{B} \rangle =
\langle \tilde{\Psi} | \hat{B} | \tilde{\Psi} \rangle - \langle \Psi_0 |
\hat{B} | \Psi_0 \rangle , \] where $\tilde{\Psi} = \Psi_0 + \delta \Psi(t)$
is the perturbed time-evolved ground state. If we limit ourselves to study
terms which are linear in the response, it is enough to consider the single
oscillatory perturbation: \[ \hat{H}'(\omega) = \frac12 (\hat{A} \emi{\omega
t} + \hat{A}^\dagger \ei{\omega t}) , \label{hpert} \] whose response can be
written, using the compact notations due to
Zubarev \cite{Zubarev60,Zubarev,McWeeny}, as: \[ \delta \langle \hat{B}
\rangle = \frac12 ( \, \langle\langle \hat{B} |\hat{A} \rangle\rangle_\omega
\emi{\omega t} + \langle\langle \hat{B}| \hat{A} \rangle\rangle_{-\omega}
\ei{\omega t} \,) .  \label{resp} \] The quantity $\langle\langle \hat{B} |
\hat{A} \rangle\rangle_\omega$ is by definition the linear response induced
by the perturbation $\hat{A}$ at frequency $\omega$ on the expectation value
$\langle \hat{B} \rangle$. Straightforward first-order perturbation theory
provides its explicit expression as: \bea \langle\langle \hat{B} | \hat{A}
\rangle\rangle_\omega &=& \frac{1}{\hbar} \lim_{\eta \rightarrow 0+} {\sum_{n
\neq 0}}' \left( \frac{\langle \Psi_0 | \hat{B} | \Psi_n \rangle \langle
\Psi_n | \hat{A} | \Psi_0 \rangle }{\omega - \omega_{0n} + i \eta} \right. \nn &-& \left.
\frac{\langle \Psi_0 | \hat{A} | \Psi_n \rangle \langle \Psi_n | \hat{B} |
\Psi_0 \rangle }{\omega + \omega_{0n} + i \eta} \right) , \label{kubo} \eea
where $\omega_{0n}$ are the excitation frequencies of the
unperturbed system, and the positive infinitesimal $\eta$ ensures causality. 
Expressions of the kind of \equ{kubo} go under the name of Kubo formulae.

\subsection{Kubo formula for conductivity}

The conductivity tensor $\sigma_{\alpha\beta}(\omega)$ measures the current linearly induced by an electric field: $j_\alpha = \sigma_{\alpha\beta} {\cal E}_\beta$. We therefore identify $\hat{A}$  with the potential of an electric field along $\beta$, i.e. $\hat{A} = e {\cal E} \hat{r_\beta}$, and $\hat{B}$ with the current operator $-e \hat{v}_\alpha/L^3$. An important detail must be stressed at this point.
The macroscopic field inside the sample includes by definition screening effects due to the electronic system, while the perturbation $\delta \hat{H}$ entering \equ{sch}---via the $\hat{A}$ operator---is the ``bare'', or unscreened one. This point will be discussed below (Sect. \ref{sec:shape}); for the time being we simply identify screened and unscreened fields.

The Kubo formula for conductivity is therefore \[ \sigma_{\alpha\beta}(\omega) =
-\frac{e^2}{L^3} \langle\langle \hat{v}_\alpha | \hat{r}_\beta \rangle\rangle_\omega ; \]
this is correct within OBCs, but meaningless within PBCs, owing to the explicit presence of the position operator. This, however, makes no harm, since only its off-diagonal matrix elements are required: see \equ{kubo}. As usual, we may exploit the identity $\ev{\Psi_0 | \hat{\r} | \Psi_n}  = i \ev{\Psi_0 | \hat{\v} | \Psi_n} / \omega_{0n}$. The Kubo formula becomes then \bea \sigma_{\alpha\beta}(\omega) &=& \frac{i e^2}{\hbar L^3} \lim_{\eta \rightarrow 0+} {\sum_{n
\neq 0}}'  \frac{1}{\omega_{0n}} \left( \frac{\langle \Psi_0 | \hat{v}_\alpha | \Psi_n \rangle \langle
\Psi_n | \hat{v}_\beta | \Psi_0 \rangle }{\omega - \omega_{0n} + i \eta} \right. \nn &+& \left.
\frac{\langle \Psi_0 | \hat{v}_\beta | \Psi_n \rangle \langle \Psi_n | \hat{v}_\alpha |
\Psi_0 \rangle }{\omega + \omega_{0n} + i \eta} \right) . \label{kubo2} \eea

We introduce a compact notation for  the real and imaginary parts of the numerators in \equ{kubo2}, i.e. 
 \bea \rab &=& \mbox{Re } \langle \Psi_0 | \hat{v}_\alpha | \Psi_n
\rangle \langle \Psi_n | \hat{v}_\beta | \Psi_0 \rangle  , \\ \iab  &=&
\mbox{Im } \langle \Psi_0 | \hat{v}_\alpha | \Psi_n \rangle \langle \Psi_n |
\hat{v}_\beta | \Psi_0 \rangle , \eea which are symmetric and antisymmetric, respectively. Using then \[ \lim_{\eta \rightarrow 0+} \frac{1}{x+ i \eta} = {\cal P} \frac{1}{x} -i \pi \delta(x), \]
and omitting the principal part, we separate for $\omega > 0$ the symmetric and antisymmetric parts in the conductivity tensor as \bea \mbox{Re }
\sigma_{\alpha\beta}^{(+)} (\omega) &=&\frac{\pi e^2}{\hbar L^3} {\sum_{n
\neq 0}}'   \frac{\rab }{\omega_{0n}} \delta(\omega - \omega_{0n})  \nn \mbox{Re }
\sigma_{\alpha\beta}^{(-)} (\omega) &=& = \frac{2 e^2}{\hbar
L^3} {\sum_{n \neq 0}}' \frac{\iab }{ \omega_{0n}^2 -
\omega^2}  . \label{all} \eea 

\subsection{Sum rules}

At this point, we are ready to compare with the sum-over-states formulae for the tensor $\lt$. In the present notations, we rewrite \equ{pertu4} as
\bea \mbox{Re } \lt &=& \frac{1}{N} {\sum_{n
\neq 0}}'  \frac{\rab }{\omega_{0n}^2} \nn
\mbox{Im } \lt &=&  \frac{1}{N} {\sum_{n \neq 0}}' \frac{\iab }{ \omega_{0n}^2} .
\eea A glance at \equ{all} shows that 
\bea \mbox{Re } \lt &=& \frac{\hbar L^3}{\pi e^2 N} \int_0^\infty \frac{d \omega}{\omega} \; \mbox{Re }
\sigma_{\alpha\beta}^{(+)} (\omega) \label{swm1} \\
\mbox{Im } \lt &=&   \frac{\hbar L^3}{2 e^2 N}  \mbox{Re }
\sigma_{\alpha\beta}^{(-)} (0) \label{swm2}. \eea
\equ{swm1} has been arrived at by SWM in 2000 \cite{Souza00}, and \equ{swm2} by Resta in 2005 \cite{rap127}. First of all, these identities show  that $\lt$, defined here as a
basic geometric feature, is indeed a measurable quantity (whenever it does not diverge).

As emphasized throughout this work $\lt$ is a ground-state property, while the
rhs of \eqs{swm1}{swm2} are properties of the system {\it excitations}, owing to the Kubo formula. Indeed, both \eqs{swm1}{swm2}
look like the zero-temperature
limit of a fluctuation-dissipation theorem, several forms of which are known
in statistical physics \cite{Kubo2,Forster}: in the lhs we have a ground-state fluctuation---see in particular \equ{eta3}---while the ingredient of the rhs is conductivity (dissipation).

\subsection{Screened vs. unscreened field} \label{sec:shape}

The Kubo formula for conductivity has been obtained identifying the $\hat{A}$ operator with ${\cal E} \hat{r_\beta}$, where ${\cal E}$ is the macroscopic field inside the sample. In general this is not quite correct, since instead $\hat{A} = e {\cal E}_0 \hat{r}_\beta$, where ${\cal E}_0$ is the ``bare'' field, i.e. the field that would be present inside the sample {\it in absence of screening}. The latter originates from the two-body (electron-electron) terms in the potential $\hat{V}$ entering Schr\"odinger equation. The relationship between ${\cal E}$ and ${\cal E}_0$ is not a bulk property, and depends on the shape of the sample. Alternatively, it depends on the boundary conditions assumed for integrating Poisson equation: we refer to Ref. \cite{rap_a30} for a thorough discussion. Whenever ${\cal E} \neq {\cal E}_0$, the sum rule in \equ{swm1} must be modified. 

The ground-state fluctuations, as e.g. \equ{eta3}, are fluctuations of the macroscopic polarization, which in a finite sample
induce a surface charge at the boundary. This in turn generates a depolarizing field, which counteracts polarization. Therefore the localization tensor $\lt$ depends on the sample shape, or equivalently on the boundary conditions assumed when taking the thermodynamic limit.
The choice of PBCs in \equ{sch}, however, implies ${\cal E} = {\cal E}_0$ \cite{rap131}; hence \eqs{swm1}{swm2} are correct as they stand within PBCs. This  no longer holds within OBCs: in this case \equ{swm1} needs to be modified,
while $\mbox{Im } \lt = 0$.

Ideally the equality ${\cal E} = {\cal E}_0$ corresponds to choosing a sample in the form of a slab, and to addressing the component of the fluctuation tensor $\lt$ parallel to the slab \cite{rap_a30}; the thermodynamic limit amounts then to the infinite slab thickness. Owing to the long range of Coulomb interaction the order of the limits (first a slab, then its infinite thickness) is crucial. For instance, if the limit is taken instead by considering spherical clusters of increasing radius, the SWM fluctuation-dissipation sum rule, \equ{swm1}, assumes a different form: this is discussed in Refs. \cite{rap132,rap131}. The explicit form of the generalized sum rule is given therein.

Last but not least, the effect leading to ${\cal E} \neq {\cal E}_0$ within OBCs is a pure correlation effect. It originates from explicitly correlated wavefunctions, and does not occur within mean-field theories (Hartree-Fock and Kohn-Sham) \cite{rap132,rap131}. Within such theories, therefore, the sum rules hold in the simple form of \eqs{swm1}{swm2}; the conductivity therein is the independent-particle conductivity (``uncoupled'' response in quantum-chemistry jargon).

\section{Localization in the insulating state} \label{sec:loca}

The basic tenet of the modern theory of the insulating state  is that the localization tensor $\lt$ is the ground-state property which sharply discriminates---in the spirit of Kohn's seminal work \cite{Kohn64,Kohn68}---between insulators and metals. The real part of $\lt$ remains finite in the thermodynamic limit in any insulator, while it diverges in any metal.

The theory is very general, and has found applications to various different kinds of  insulators: band
insulators \cite{rap118,Veithen02,Hine07,Monari08};  correlated (i.e.
Mott) insulators, either by means of Hubbard-like model
Hamiltonians \cite{rap107,Aebischer01} or realistic
ones \cite{Vetere08}; ``quantum Hall
insulators'' \cite{rap127}, and  Anderson insulators \cite{rap143}. As for Chern insulators and topological insulators, no explicit application of the present theory exists; nonetheless the work of Refs. \cite{Thonhauser06,Z2}
implicitly shows that even in these cases the ground-state wavefunction is indeed localized in the sense of the present review. Most of these applications are reviewed in Sect. \ref{sec:kinds}.

The ultimate proof of the key property of $\mbox{Re } \lt$ is based on the SWM sum rule, \equ{swm1}. Since the tensor is real symmetric, it is enough to consider the diagonal elements (over its principal axes) \[ \mbox{Re } \langle r_\alpha r_\alpha \rangle_{\rm c}  = \frac{\hbar L^3}{\pi e^2 N} \int_0^\infty \frac{d \omega}{\omega} \; \mbox{Re }\sigma_{\alpha\alpha} (\omega) . \label{swm} \]
The $f$-sum rule yields \[ \int_0^\infty d \omega \; \mbox{Re } \sigma_{\alpha\alpha} (\omega) = \frac{\omega_{\rm p}^2}{8} = \frac{\pi e^2 N}{2 m_e L^3} , \label{fsum} \] where $\omega_{\rm p}$ is the plasma frequency. Therefore the integral in \equ{swm} always converges at $\infty$; its convergence/divergence is dominated by the small-$\omega$ behavior of $\mbox{Re } \sigma_{\alpha\alpha} (\omega)$.

Suppose first that the spectrum is gapped, i.e. the spacing between the ground state and the first excited state stays finite in the thermodynamic limit. If the gap is $E_{\rm g}$ the conductivity vanishes for $\omega < E_{\rm g}/\hbar$, and \equ{fsum} yields \bea \mbox{Re } \langle r_\alpha r_\alpha \rangle_{\rm c}  &=& \frac{\hbar L^3}{\pi e^2 N} \int_{E_{\rm g}/\hbar} ^\infty \frac{d \omega}{\omega} \; \mbox{Re }\sigma_{\alpha\alpha} (\omega) \nn &<& \frac{\hbar^2 L^3}{\pi e^2 N E_{\rm g}} \int_{0} ^\infty d \omega \; \mbox{Re }\sigma_{\alpha\alpha} (\omega)  \nn &=& \frac{\hbar^2}{2 m_e E_{\rm g}} . \label{maj} \eea This inequality is due to SWM and clearly proves that $\mbox{Re } \lt$ is finite in any gapped insulator, as e.g. band insulators (considered in more detail in Sect. \ref{sec:bands}).

The main message of Kohn's 1964 paper, however, is that ``insulating characteristics are a strict consequence of electronic localization (in an appropriate sense) and do not require an energy gap''. For any gapless material, the small-$\omega$ behavior of $\mbox{Re } \sigma_{\alpha\alpha} (\omega)$ is the result of a competition between numerators and denominators in the Kubo formula, \equ{kubo2}.  Since we aim at a continuous function of $\omega$, the singularities in \equ{all} must be smoothed: this can be done by keeping the ``dissipation'' $\eta$ finite while performing the thermodynamic limit
first~\cite{Akkermans97}. For a band metal the localization tensor diverges (see below). According to SWM, a gapless material is insulating whenever $\mbox{Re } \sigma_{\alpha\alpha} (\omega) \rightarrow 0$ like a positive power of $\omega$,
and metallic otherwise. The only example of gapless insulator considered so far is a model Anderson insulator in 1d \cite{rap143}. Simulations prove indeed that $\langle x^2 \rangle_{\rm c}$ is finite therein (Sect. \ref{sec:anderson}).

\subsection{Independent electrons}

For noninteracting electrons the potential $\hat{V}$ in \equ{sch} is the sum of identical one-body terms: $\hat{V} = \sum_{i=1}^N V(\r_i)$. The many-electron Hamiltonian is separable and the exact ground state $\ket{\Psi_0(\kk)}$ is a Slater determinant of one-particle orbitals (doubly occupied in the singlet case). At a mean-field level, the one-body potential $V(\r)$ includes electron-electron interaction in a selfconsistent way. In the Hartree-Fock (HF) framework the Slater determinant is regarded as an approximate many-electron wavefunction. Instead, in the density-functional framework the orbitals---called Kohn-Sham (KS) orbitals---are auxiliary quantities, individually devoid of  physical meaning. In particular, their Slater determinant {\it does not} coincide with the many-electron wavefunction
$\ket{\Psi_0(\kk)}$, as a matter of principle. Therefore the exact localization tensor 
does not coincide, at least in principle, with the one obtained from the Slater determinant of KS orbitals.

So much for the matters of principle. On practical grounds such difference is routinely disregarded (e.g. when dealing with polarization, magnetization  \cite{rap_a27,rap_a28,rap_a30}, and more), given that it is not at all clear what is the relative importance of this ``intrinsic'' error, compared with the errors due to the choice of the functional itself. We therefore address here either the HF or the KS wavefunction $\ket{\Psi_0(\kk)}$, having the form of a Slater determinant.

Whenever the wavefunction is a Slater determinant, all ground-state properties 
can be explicitly cast in terms of the one-body density matrix  \[ \rho({\bf r},{\bf
r}') = 2 P(\r,\r') = 2 \sum_{j=1}^{N/2} \varphi_j({\bf r})
\varphi_j^*({\bf r}') , \label{rho} \] where a singlet ground state is assumed, and $\varphi_j({\bf r})$ are the occupied one-particle orbitals (either HF or KS); $P({\bf r},{\bf r}')$ is  the projector over the occupied manifold.

The expression  for the localization tensor is easily found within OBCs starting from \equ{eta3} \cite{rap132}: \[ \lt = \frac{1}{N} \int
d {\bf r} d {\bf r}'\;({\bf r} - {\bf r}')_\alpha ({\bf r} - {\bf r}')_\beta \,
|P(\r,\r')|^2 . \label{cform} \]  If we define the complementary projector \[ Q(\r,\r') = \delta(\r - \r') - P(\r,\r') , \] an equivalent expression is \cite{rap118} \[ \lt = \frac{2}{N}
\mbox{Tr } \{ r_\alpha P r_\beta Q \} , \] where ``Tr'' is the trace over the single-particle Hilbert space (not on Cartesian indices).

If we consider a cluster, cut out of a crystalline solid, \equ{cform} becomes in the large-$N$ limit \begin{equation} \langle r_\alpha r_\beta \rangle_{\rm
c} = \frac{1}{N_{\rm c}} \int_{\mbox{\scriptsize cell}} \!\!\! d {\bf r}
\int_{\mbox{\scriptsize all space}} \!\!\!\!\!\!\!\!\!\!\!\! d {\bf r}' \; ({\bf
r - r'})_\alpha ({\bf r - r'})_\beta \, |P({\bf r}, {\bf r}')|^2 \label{dform} ,
\end{equation} where $N_{\rm c}$ is the number of electrons per crystal cell. According to the discussion in Sec. \ref{sec:shape}, we need not to worry about shape
issues in taking the limit; we also notice that the density matrix, \equ{rho}, is independent of the boundary conditions (either OBCs or PBCs) in the large-$N$ limit. 

\subsection{Band insulators and band metals} \label{sec:bands}

As observed, \equ{dform} holds for a crystalline solids. Therefore the inner integral on the rhs must converge for a band insulator, and must diverge for a band metal.
This is confirmed  by the well known fact that the asymptotic behavior of $P$ is qualitatively different in
insulators and in metals. In the former materials, in fact, $P({\bf r}, {\bf
r}')$ decays exponentially \cite{Kohn59,desCloizeaux64,Ismail99,He01} for large
values of ${\bf r} - {\bf r}'$: therefore the integral
converges and the localization tensor is finite. In conducting materials,
instead, $P({\bf r}, {\bf r}')$ decays only polynominially, and the inner
integral diverges. This divergence can be explicitly verified for the simplest
conductor of all, namely, the noninteracting electron gas, whose density matrix
is exactly known in analytic form \cite{rap118,GiulianiVignale}. Therefore the localization tensor, when expressed in the form of  \equ{dform}, measures in a perspicuous way the ``nearsightedness''\cite{Kohn96} of the electron distribution. Such measure is {\it qualitatively} different in insulators and in metals.

The one-particle orbitals (either HF or KS) in a crystalline solid have the Bloch
form. We therefore may wish to replace the orbitals in the expression for $P(\r,\r')$, \equ{rho}, \[ \varphi_i(\r) \rightarrow \psi_{n\q}(\r) = \ei{\q \cdot \r} u_{n\q}(\r) , \] where $n$ is the band index and $\q$ is the Bloch vector. We stress that PBCs are at the very root of Bloch theorem. If the orbitals are normalized to one over the crystal cell of volume $V_{\rm c}$, the ground-state projector in insulating crystals is
\bea P(\r,\r') &=& \frac{V_{\rm c}}{(2\pi)^3} \sum_{n=1}^{N_{\rm c}/2} \intq \psi_{n\q}(\r) \psi_{n\q}^*(\r') \label{rhob} \\ &=& \frac{V_{\rm c}}{(2\pi)^3} \sum_{n=1}^{N_{\rm c}/2} \intq \ei{\q \cdot (\r - \r')} u_{n\q}(\r) u_{n\q}^*(\r') , \nonumber \eea
where $N_{\rm c}/2$ is the number of occupied bands, and the integral is taken over the Brillouin zone (equivalently, over the reciprocal cell).

Using \equ{rhob}, the localization tensor in a band insulator becomes  (for double occupancy) \bea  \langle r_\alpha
r_\beta \rangle_{\rm c} &=& \frac{2 V_c}{(2\pi)^3 N_{\rm c}} \intq
\left( \sum_n \langle \frac{\partial}{\partial q_\alpha} u_{n{\bf q}} |
\frac{\partial}{\partial q_\beta} u_{n{\bf q}} \rangle \right. \nonumber \\ &&
\left.  - \sum_{n,n'} \langle u_{n{\bf q}} | \frac{\partial}{\partial
q_\alpha} u_{n'{\bf q}} \rangle \langle \frac{\partial}{\partial q_\beta}
u_{n'{\bf q}} | u_{n{\bf q}} \rangle \right) . \label{mom2} \eea The proof is given in Refs. \cite{rap_a23,rap118}, and will not be repeated here.

\subsection{Wannier functions}

Expressions similar to \equ{mom2} also enter the Marzari-Vanderbilt theory of maximally localized Wannier functions \cite{Marzari97}; the relationship is \[
\sum_{\alpha=1}^{d}  \langle r_\alpha
r_\alpha \rangle_{\rm c} = \frac{2}{N_{\rm c}} \Omega_{\rm I} . \label{Om} \] Here $\Omega_{\rm I}$ indicates the gauge-invariant part of the quadratic spread of the Wannier functions, as in Ref. \cite{Marzari97} and in the subsequent literature.

The trace in \equ{Om} is a lower bound (and {\it not} a minimum in dimension d $>$ 1) for the spherical second (cumulant) moment---a.k.a. quadratic spread---of the Wannier functions, averaged over the sample.

We stress that \eqs{rhob}{mom2} make sense only insofar the Fermi level falls in a gap, in which case $\lt$ is always finite. If we vary the Hamiltonian continuously, allowing the gap to close, then $\lt$ diverges; the quadratic spread of the Wannier functions diverges as well \cite{rap132}.

\subsection{Chern insulators} \label{sec:cins}

Within OBCs the localization tensor is always real: this is perspicuous in \eqs{cform}{dform}. Instead the PBC expression of \equ{mom2} is naturally endowed with an imaginary part: \bea \mbox{Im } \langle r_\alpha
r_\beta \rangle_{\rm c} &=& \frac{2 V_c}{(2\pi)^3 N_{\rm c}} \times \nn &\times&\intq \sum_n \mbox{Im } \langle \frac{\partial}{\partial q_\alpha} u_{n{\bf q}} |
\frac{\partial}{\partial q_\beta} u_{n{\bf q}} \rangle . \label{ima} \eea A necessary condition for this to be nonzero is the absence of time-reversal symmetry.

The second line of \equ{ima} has (in 3d) the dimensions of an inverse length, and is quantized in units of $\pi$ times a reciprocal vector. We call this reciprocal vector ``Chern invariant''; in fact it is a Chern number (Sect. \ref{sec:chern}) in $\q$-space and in appropriate units. In 2d the BZ integral  is dimensionless and 
simply related to the Chern number, defined as:  
\[  C_1 =  - \frac{1}{\pi} \intq \sum_n \mbox{Im } \langle \frac{\partial}{\partial q_1} u_{n{\bf q}} |
\frac{\partial}{\partial q_2} u_{n{\bf q}} \rangle \label{c1} \]  (the formula here is given for single band occupancy).

Band insulators where the Chern invariant is nonzero are called ``Chern insulators'' (normal insulators otherwise). The possiible existence of Chern insulators has been pointed out by Haldane in 1988 \cite{Haldane88}, by means of a remarkable model Hamiltonian in 2d, much studied afterwards \cite{rap130,Thonhauser06,Coh09}, and illustrated below (Sect. \ref{sec:c1}).
We emphasize that the nonvanishing of the Chern number prevents the existence of exponentially localized Wannier functions \cite{Thonhauser06,Thouless84}, at variance with normal insulators where the ``maximally localized'' Wannier functions \cite{Marzari97} {\it are} exponentially localized \cite{Brouder07}.

\section{Localization in different kinds of insulators} \label{sec:kinds}

\subsection{Small molecules}

The modern theory of the insulating state clearly addresses extended systems, i.e. the $N \rightarrow \infty$ limit; indeed it makes little sense to ask whether a small molecule is insulating or conducting. Nonetheless the concepts of localized/delocalized electronic states is of the utmost importance in quantum chemistry
as well, notably in relationship to aromaticity. 

The tensor $\lt$ within OBCs is always real symmetric. If the ground-state wavefunction is a Slater determinant, then the trace of the tensor at finite $N$
has the meaning of a lower bound for the quadratic spread of the Boys localized orbitals, averaged over all the occupied orbitals \cite{rap132}.

The small-$N$ version of the main concepts of the present review  (in their OBCs flavour \cite{rap132}) has been adopted  in quantum chemistry by \`Angy\`an \cite{Angyan09,Angyan10}. Besides providing HF calculations of $\lt$ for a sample of small molecules, \`Angy\`an even provides {\it experimental values} drawn from compilations of the dipole oscillation-strength
distributions: basically, from \equ{swm1}.

\subsection{Band insulators}

The theory warrants that the localization tensor is finite in any insulator. However, 
quantitative calculations for both model tight-binding Hamiltonians and realistic solids within density-functional theory have been used to illustrate the theory and to identify trends. For instance one expects much smaller diagonal elements $\langle r_\alpha r_\alpha \rangle_{\rm c}$ for strong (i.e. large-gap) insulators than for weak (small-gap) ones. This is also suggested by the SWM inequality, \equ{maj}.

Let us start with a simple tight-binding (a.k.a. H\"uckel) Hamiltonian in 1d:
\[ \hat{H} = \sum_{j \sigma} [ \; (-1)^j \Delta \,
c^\dagger_{j \sigma} c_{j \sigma} - t ( c^\dagger_{j \sigma} c_{j+1 \sigma} +
\mbox{H.c.} ) \; ] \label{tight} \] where
$t > 0$ is the first neighbor hopping ($\beta = - t$ in most chemistry
literature) and H.c. stays for Hermitian conjugate. This toy model schematizes a binary ionic crystal; the band structure is
\[ \epsilon(q) = \pm \sqrt{\Delta^2 + 4 t^2 \cos^2 qa/2} , \label{band} \] where $a$ is the lattice constant and $q$ is the Bloch vector. The gap is equal to $2 \Delta$; at half filling the system is always insulating except for $\Delta=0$.
The squared localization length (within OBCs) is the tight-binding version of \equ{cform}, i.e.
 \[ \xc = \frac{a^2}{4 N} \sum_{j,j'=1}^N P_{jj'}^2 (j-j')^2 . \label{lambda} \] 
This is a monothonical function of $t/\Delta$; it is easily verified that it vanishes in the extreme ionic case 
($t=0$). In the metallic  case ($\Delta=0$) the ground-state projector has a simple analytical form: \bea P_{jj} &=& \frac{1}{2} ;  \quad P_{jj'} = 0 \; \mbox{for even } |j' - j| = 2s, \nn P_{jj'} &=& \frac{(-1)^s}{\pi (2s + 1)}\;  \mbox{for odd } |j' - j| = 2s+1 , \label{exact} \eea which clearly implies divergence of \equ{lambda}. At any finite $N$ within OBCs \equ{lambda} leads to a finite $\xc$ value; however \equ{exact} suggests that in the metallic case $\xc$ diverges linearly with $N$. This has been verified by actual simulations, even when $\Delta \neq 0$ but the Fermi level is not in the gap \cite{rap143}. 

Other simulations \cite{Monari08} have addressed dimerized chains, i.e. $\Delta=0$ but alternant hoppings in \equ{tight}. While nothing relevant occurs within PBCs, partly filled end states within OBCs at some fillings are at the root of some noticeable features.

\begin{figure} 
\centerline{\includegraphics[width=7cm]{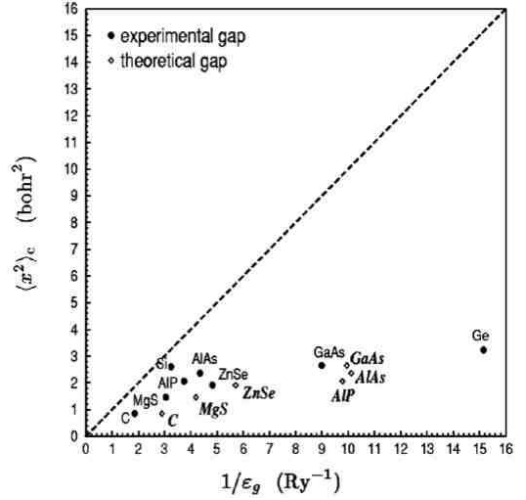}}
\caption{Diagonal element of the KS localization tensor vs. the inverse direct gap
(theoretical and experimental), for several elemental and binary
semiconductors (from Ref. \protect\cite{rap118}) The points corresponding to Si and Ge with the theoretical gaps
are out of scale. From Ref. \cite{rap118}.}
\label{fig:sgiaro} \end{figure}

The first ab-initio study (in 2001) addressed several elemental and binary cubic semiconductors at the KS level \cite{rap118}. The tensor is real and isotropic. The computed $\xc$ (Fig. \ref{fig:sgiaro}) is smaller than 3 bohr$^2$ in all the materials studied:
the ground many-body wavefunction is therefore very localized in this class of materials. The SWM inequality was also checked, and found to be well verified using both the theoretical KS gap and the experimental one (the latter is typically larger).

Other studies have addressed the ferroelectric perovskites in their different (cubic and noncubic) structures  \cite{Veithen02}, and some model Hamiltonians in 1d and 2d \cite{Hine07}.

\subsection{Correlated (Mott) insulators}

\begin{figure} 
\centerline{\includegraphics[width=8cm]{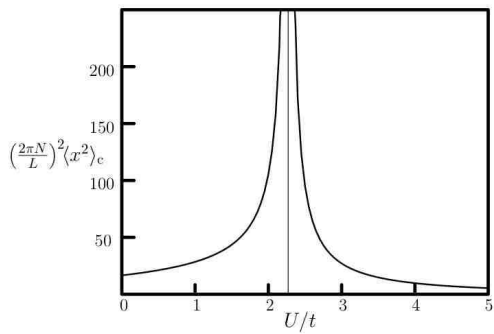}}
\caption{Squared localization length for the Hamiltonian in \protect\equ{hub} at half filling for $t/\Delta = 1.75$. The system undergoes a quantum phase transition from band-like insulator to Mott-like insulator at $U/t=2.27$. From Ref. \protect\cite{rap107}}
\label{fig:hubbard} \end{figure}

Starting from the noninteracting Hamiltonian of \equ{tight} and augmenting it with an on-site repulsive term we get the two-band Hubbard model
\[ \hat{H} \! = \! \sum_{j \sigma} [  (-1)^j \Delta \,
c^\dagger_{j \sigma} c_{j \sigma} - t ( c^\dagger_{j \sigma} c_{j+1 \sigma} +
\mbox{H.c.} ) ] + \, U \sum_j n_{j\uparrow}  n_{j\downarrow}  . \label{hub} \] 
The explicitly correlated ground-state wavefunction has been found by exact diagonalization \cite{rap107}, and the corresponding $\xc$ has been computed as a function of $U$ for fixed $t/\Delta = 1.75$. The results are shown in Fig. \ref{fig:hubbard} in dimensionless units; it turns out that there is only one singular point $U=2.27t$, where $\xc$ diverges. Indeed, it has been verified that at such value the ground-state becomes degenerate with the first excited singlet state, i.e. the system is metallic. The singular point is the fingerprint of a quantum phase transition: on the left we have a band-like insulator, and on the right a Mott-like insulator. The two insulating states are {\it qualitatively} different; by adopting the modern jargon, nowadays we could say that they are {\it topologically} distinct. The static ionic charges (on anion and cation) are continuous across the transition, while the dynamical (Born) effective charge on a given site changes sign \cite{rap87}. Other studies of the localization tensor within the same Hubbard model can be found in Ref. \cite{Wilkens01}.

The transition from a band metal to a Mott insulator has been studied in a model linear chain of Li atoms by Vetere {\it et al.} \cite{Vetere08}. At a mean-field level the infinite chain is obviously metallic at any lattice constant $a$, since there is one valence electron per cell. However the mean-field description becomes inadequate at large $a$, where the electrons localize and the system becomes a Mott insulator. 
If electron correlation is properly accounted for at any $a$, the system undergoes a sharp metal-insulator transition at a critical $a$. 

The calculations addressed linear Li$_N$ systems ($N$ up to 8), where the finite size prevents a sharp transition; the tradeoff is that full configuration interaction was affordable with 6 atomic orbitals per site (yielding more than 10$^9$ symmetry-adapted Slater determinants). The wavefunction  of Vetere {\it et al.} is therefore exempt from any bias insofar as the treatment of correlation is concerned, although its quality
is determined by the basis set. A study of the longitudinal component $\xc$ of the localization tensor indicates rather clearly the occurrence of the metal-insulator transition at $a \simeq 7$ bohr; other indicators give concordant results \cite{Vetere08}. For comparison, the nearest-neighbour distance in 3d metallic lithium is 5.73 bohr.

\subsection{Disordered (Anderson) insulators} \label{sec:anderson}

\begin{figure} 
\centerline{\includegraphics[width=8.5cm]{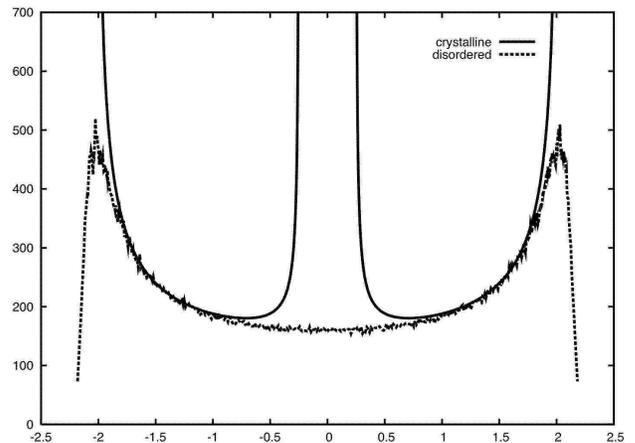}}
\caption{Density of states (arbitrary units) for a model binary alloy in 1d. The crystalline (band) case corresponds to the Hamiltonian of \protect\equ{tight} with $\Delta=0.25$ and $t=1$. The disordered (Anderson) case corresponds to a random choice of the anion/cation distribution. }
\label{fig:dos} \end{figure}

We start from the same Hamiltonian as in \equ{tight}, and we replace the ordered string  $(-1)^j$ by a random string of $\pm1$, chosen with equal (and uncorrelated) probability. This system models a random binary alloy at 50\% concentration. It is well known both from analytical arguments and actual simulations that its spectrum is gapless \cite{Abrahams,Kramer93}. The density of states for both the ordered and disordered systems are shown in Fig. \ref{fig:dos}, and confirm the expected features. The band structure of \equ{band} yields obviously a gapped density of states; at the band edges it shows van Hove singularities, which in 1d  have the character of $1/\sqrt{\epsilon}$ divergences. As discussed above, the system is insulating at half filling and conducting otherwise. The disordered system, instead, is gapless and nonetheless insulating at any filling. In fact, this model Hamiltonian describes a paradigmatic Anderson insulator in 1d.

The conventional theory of transport focusses on the nature of the one-particle orbitals at the Fermi level; in Anderson insulators these are localized, thus forbidding steady state currents \cite{Anderson58}. More than fifty years of literature have been devoted to investigate Anderson insulators under the most diverse aspects. \cite{Abrahams,Kramer93,Thouless74,Lagendijk09}. 

At variance with such wisdom, a recent work has 
addressed this paradigmatic Anderson insulator from the nonconventional viewpoint of the modern theory of the insulating state \cite{rap143}. In the spirit of Kohn's theory the invidual Hamiltonian eigenstates become apparently irrelevant, while the focus is on the many-electron ground state as a whole. The squared localization length $\xc$ has been computed within OBCs from \equ{lambda}, and found to be finite, as expected. Nonetheless its value is about 20 times larger than the one for the band insulator, at the same value of the parameters (i.e. $\Delta = 0.25, t=1$). This reflects the fact that the scattering mechanisms are profoundly different: incoherent (Anderson) versus coherent (band). In the latter case, the Hamiltonian eigenstates are individually conducting but``locked'' by the Pauli principle if the Fermi level lies in the gap. 

\subsection{Quantum Hall insulators} \label{sec:qh}

\begin{figure} 
\centerline{\includegraphics[width=7cm]{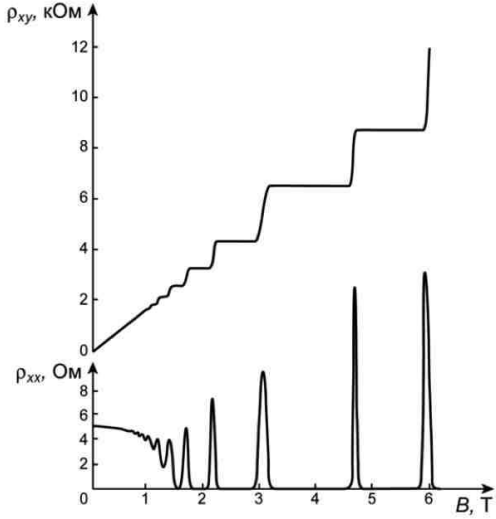}}
\caption{The results of a typical quantum Hall experiment.}
\label{fig:qhe} \end{figure}

The results of a typical quantum Hall (QH) experiment are shown in Fig. \ref{fig:qhe}. The most perspicuous feature is that whenever the 2d electron fluid is in the QH regime: (i) the transverse resistivity is quantized; and (ii) the longitudinal resistivity vanishes. While the plots are about the integer QH effect, the same two features are common to the fractional QH effect as well. 

We switch from resistivity to conductivity; in isotropic 2d-systems the relationship is
\[ \rho_{11} = \frac{\sigma_{11}}{\sigma_{11}^2 + \sigma_{12}^2} \qquad
\rho_{12} = - \frac{\sigma_{12}}{\sigma_{11}^2 + \sigma_{12}^2} , \] hence in he QH regime $\sigma_{11} = 0$ and $\sigma_{12} = - 1/ \rho_{12}$ is quantized.
Since the static longitudinal conductivity vanishes, the system is by definition insulating: this is why we speak of ``QH insulators''.

We are going to show that electron localization---as defined in the modern theory of the insulating state---is the common {\it cause} for both the vanishing of the dc longitudinal conductivity and the quantization of the transverse one. Therefore to predict whether the dc transverse conductivity is quantized, it is enough to inspect
electron localization in the ground state. This outstanding result was proved in 2005 \cite{rap127}.

Our system is a square of size $L \times L$, where the presence of a macroscopic $B$ field {\it forbids}
simple PBCs; this, however, is no serious problem. We choose the Landau gauge for the magnetic vector potential in \equ{sch}, and we impose the magnetic boundary conditions which are customary in the QH literature \cite{Niu85}; this requires the total flux $BL^2$ to be an integer times $\Phi_0$, where $\Phi_0 = hc/e$ is the flux quantum.

When switching from 3d to 2d, many of the above formulae require obvious modifications. For an isotropic system, the 2d analogue of \eqs{swm1}{swm2} can be written as \bea \int_0^\infty \frac{d
\omega}{\omega} \; \mbox{Re }  \sigma_{11}(\omega) & = &
\frac{\pi e^2 N}{\hbar L^2} \mbox{Re } \ev{x^2}_{\rm c} , \label{sum1} \\ \mbox{Re } \sigma_{12}(0) & = & \frac{2 e^2 N}{\hbar L^2} \mbox{Im } \ev{xy}_{\rm c} .
\label{sum2} \eea A glance at the sum-over-states expression for $\lt$, \equ{pertu4}, shows that whenever $ \mbox{Re } \lt$ converges, then $ \mbox{Im } \lt$
converges as well. However, we are going to show that the latter may only converge to quantized values in the present 2d case.

By definition we have \[ \mbox{Im } \ev{xy}_{\rm c} = \frac{1}{N} \mbox{Im } \ev{\partial_1 \Psi_0(0) | \partial_2 \Psi_0(0)} . \] In the large-$L$ limit, we may write
\bea \mbox{Im} && \!\!\!\! \ev{\partial_1 \Psi_0(0) | \partial_2 \Psi_0(0)}\\ &= &\left( \frac{L}{2\pi} \right)^2 \mbox{Im} \int_0^{\frac{2\pi}{L}} \!\!\!\! d \kappa_1 \int_0^{\frac{2\pi}{L}} \!\!\!\! d \kappa_2 \; \ev{\partial_1 \Psi_0(\kk) | \partial_2 \Psi_0(\kk)} . \nonumber \eea The dimensionless integral is over a closed surface (a torus), given the magnetic boundary conditions. It is therefore equal to $-\pi C_1$, where $C_1$ is the Chern number of the first class (Sect. \ref{sec:chern});
hence \[ \mbox{Im } \ev{xy}_{\rm c} = - \frac{L^2}{4 \pi N} C_1 , \label{xy} \] and the transverse conductivity, \equ{sum2} becomes
 \[ \mbox{Re } \sigma_{12}(0) = - \frac{e^2}{2 \pi \hbar} C_1 = - \frac{e^2}{h} C_1 \label{ntw} . \]  We have thus arrived at the famous Niu-Thouless-Wu formula \cite{Niu85} which holds for both the integer and fractional QH effect. The original derivation was based on an analysis of the Green function, under the hypothesis that the system has a Fermi gap; in the
present derivation the presence of a Fermi gap becomes apparently irrelevant, since the localization tensor is a pure {\it ground-state} property. Underlying the modern theory of the insulating state, however, is a fluctuation-dissipation theorem, relating the ground state to the excitations of the system.

The main message of Ref. \cite{rap127} (and of the present Section) can be 
summarized by saying that {\it any} 2d insulator displays a quantized transverse conductance (nonvanishing only in absence of time-reversal symmetry).

An electron fluid is kept in the QH regime by disorder, and analytical implementations of the present formulae are obviously not possible. Nonetheless, in order to illustrate how the theory works, it is expedient to consider the academic case of noninteracting electrons in a flat substrate potential. If the first Landau level is fully occupied (i.e. at filling $\nu = 1$) the electron density is uniform and equal to $n_0 = 1/(2
\pi \ell^2)$,  where $\ell = (\hbar c/e B)^{1/2}$ is the magnetic
length; the modulus of the density matrix has the simple expression (for single occupancy) \[ |\rho(\r,\r')| = |P(\r,\r')| = n_0 {\rm e}^{-|\r - \r'|^2/(4 \ell^2)} , \label{gauss} \] and the cumulant second moment is clearly finite. The trace $\langle x^2 \rangle_{\rm c} + \langle y^2 \rangle_{\rm c}$ of the localization tensor is in fact equal to $\ell^2$, the squared magnetic length. If the filling is fractional the density cannot be uniform (for noninteracting electrons), and, more important, the cumulant second moment diverges (see Ref. \cite{rap127} for details): the system is not insulating, and its transverse conductivity is not quantized.

\subsection{Chern insulators} \label{sec:c1}

The QH state of matter, discovered in 1980 and addressed in the previous Section, provides the first example of a quantum state which is topologically distinct  from all previously known states of matter. This is the reason why its macroscopic quantization properties are very robust (``topologically protected''), and insensitive to small changes in material parameters.
The  debut of geometrical concepts in electronic structure theory dates from two milestone papers: TKNN in 1982 \cite{Thouless82} and Niu-Thouless-Wu in 1985 \cite{Niu85}. These papers addressed a 2d electron fluid (noninteracting and interacting, respectively) in the presence of a macroscopic magnetic field: the Hamiltonian, therefore, cannot be lattice periodical.

A subsequent breakthrough on the theory side is the Haldane model Hamiltonian~\cite{Haldane88}: this can be considered as the precursor of
modern topological insulators \cite{Hasan10}. Its trademark is quantized Hall conductance in absence of a macroscopic magnetic field.

\begin{figure}
\centerline{\includegraphics[width=7cm]{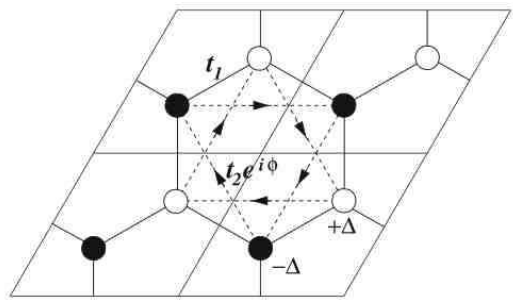}}
\caption{Four unit cells of the Haldane model~\cite{Haldane88}.
  Filled (open) circles denote sites with $E_0=-\Delta$ ($+\Delta$).
  Solid lines connecting nearest neighbors indicate a real hopping
  amplitude $t_1$; dashed arrows pointing to a second-neighbor site
  indicates a complex hopping amplitude $t_2e^{i\phi}$.
  Arrows indicate sign of the phase $\phi$ for second-neighbor hopping.
} \label{fig:haldane} \end{figure}

The model is
comprised of a 2d  honeycomb lattice with two tight-binding sites per primitive
cell with site energies $\pm \Delta$, real first-neighbor hopping $t_1$, and
complex second-neighbor hopping $t_2e^{\pm i\phi}$, as shown in
Fig.~\ref{fig:haldane}. Within this two-band model, one deals with insulators by
taking the lowest band as occupied. The appeal of the model is that the vector potential and the Hamiltonian are lattice periodical and the single-particle orbitals always have the usual Bloch form.  Essentially, the microscopic magnetic field can be thought as staggered (i.e. up and down in different regions of the cell), but its cell average vanishes.

For a 2d lattice-periodical Hamiltonian the Chern number $C_1$ has been defined before, \equ{c1}.
As a function of the flux parameter $\phi$, this system undergoes
a transition from a normal insulator ($C_1=0$) to Chern insulator ($|C_1|=1$). 

Like in the QH case discussed in Sect. \ref{sec:qh}: (i) the imaginary part of the localization tensor is related to the Chern number by \equ{xy}; and (ii) the trace of the localization tensor is finite. This is confirmed by
 the simulations of Ref. \cite{Thonhauser06}, where the actual value of $\Omega_{\rm I}$, \equ{Om}, is computed. We remind that, despite $\Omega_{\rm I}$ being finite, localized Wannier functions (with finite quadratic spread) do not exist in Chern insulators.
 
The Haldane model insulator and the QH insulator both display quantized transverse conductivity, and both are localized in the sense of the modern theory of the insulating state. It is worth pointing out, though, that the decay of the density matrix is qualitatively different: exponential in the Haldane case \cite{Thonhauser06} vs. Gaussian in the (noninteracting) QH case, \equ{gauss}.

No microscopic  realization of a Chern insulator (in absence of a macroscopic magnetic field) is known to date. A {\it mesoscopic} Chern insulator, in the same spirit as the Haldane model, was synthesized in 2008 \cite{Taillefumier08}.
Chern insulators remained a curiosity of academic interest only for many years. In 2005 it was realized that the Haldane model is the forebear of a completely new class of insulators, called ``topological insulators'' \cite{Kane05,Hasan10}.

\subsection{Topological insulators} \label{sec:topo}

Although Chern insulators do not exist in nature, a different kind of topological insulators {\it does exist} in nature  in both 2d and 3d \cite{Sheng06,Zhang08,Chen09,Moore09,Qi10,Hasan10}; 
their macroscopic quantization properties are topologically protected, and insensitive to small changes in material parameters, exactly like in the QH case.
 
Modern topological insulators are characterized by novel invariants  called \ZZ \ (close relatives of the Chern number in 2d and of the Chern invariant in 3d).
The paradigmatic \ZZ \ topological insulator is the Kane-Mele model Hamiltonian
in 2d \cite{Kane05}. Most important, at variance with Chern insulators, the modern topological insulators  are time-reversal symmetric. 

Since spin-orbit interaction plays a major role, the definition of localization tensor needs to be augmented to explicitly include the spin coordinates; this seems to be rather straightforward, and is deferred to further publications. After this is done,
the scope of the modern theory of the insulating state would include topological insulators.

Obviously, no investigation of this kind exists yet. Nonetheless, a closely related investigation concerning the Wannier functions in a \ZZ \ insulator is appearing these days \cite{Z2} . The results confirm that the ground state of a topological insulator is localized in the sense of the modern theory of the insulating state: $\Omega_{\rm I}$, \equ{Om}, is in fact finite. Furthermore---at variance with the Chern-insulator case---localized Wannier functions (with finite quadratic spread) do exist, although a generalization of the Marzari-Vanderbilt  theory \cite{Z2,Marzari97} is needed.

\section{Conclusions}

According to the milestone 1964 Kohn's paper \cite{Kohn64}, the organization of the electrons in their {\it ground state} is qualitatively different in insulators and metals. This paper provided the foundations of the ``theory of the insulating state''. The modern developments of the theory  started many years later (1999) with the work of Resta and Sorella  \cite{rap107}, and continue to these days.

Here we provide a comprehensive presentation of the modern theory. In so doing, we emphasize the geometrical properties of the many-body wavefunction, which were somewhat hidden in most of the original literature. 

We focus on the linear response of the many-body wavefunction to an infinitesimal ``twist'' (or ``flux'') in the electronic Hamiltonian. This is the ground-state property which discriminates insulators from metals. The distance between the (infinitesimally) twisted and untwisted ground states, measured by the quantum metric tensor in the thermodynamic limit, is finite in insulators and divergent in metals.

Technically, in our formulation the twist has the dimensions of an inverse length; the quantum metric tensor has therefore the dimensions of a squared length and is extensive. The key intensive quantity is the metric tensor {\it per electron}; this is generally called ``localization tensor'' or ``second cumulant moment'' of the electron distribution. It is a global property of the electronic ground state, apparently unrelated to the system excitations. In some cases (discussed above) the localization tensor is naturally endowed with an imaginary antisymmetric part, whose physical meaning is also outstanding.

The modern theory of the insulating state encompasses all kinds of known insulators. Here we discuss, using uniform concepts and notations, the most diverse cases: open and periodic boundary conditions; interacting and noninteracting electrons, crystalline and disordered systems; time-reversal symmetry present and absent. We even briefly comment on topological insulators.

Work partly supported by the ONR grant N00014-07-1-1095.



\end{document}